\documentclass[11pt,a4paper]{article}
\usepackage{jheppub}
\usepackage[utf8]{inputenc}
\usepackage{graphicx,fancybox,float,comment,bm}

\usepackage[dvipsnames]{xcolor}
\usepackage{amsmath,amssymb,amsfonts}
\usepackage{ulem}

\newcommand{\AY}[1]{{\textcolor{Green}{ [AY:#1] }}}

\title{Long lived quasi normal modes}
\author[a,b]{Sebastian Waeber}
\author[b]{Amos Yarom}

\affiliation[a]{Department of Physics, Ben-Gurion University of the Negev,
David Ben Gurion Boulevard 1, Beer Sheva 84105, Israel}
\affiliation[b]{Department of Physics, Technion, Haifa 32000, Israel}
\emailAdd{s.f.d.waeber@gmail.com}
\emailAdd{ayarom@physics.technion.ac.il}

\abstract{We consider magnetically charged AdS black branes with vanishing entropy at zero temperature. We argue that in the presence of a large enough Chern-Simons coupling the quasi normal modes of the brane will have a diminishing imaginary  part. Since our result is agnostic to the matter content of the theory it implies that, generically, the boundary theory will possess long lived modes which should be observable in a proper setting. }

\begin{document}
\maketitle

\section{Introduction}

The gauge gravity duality provides a powerful tool in uncovering novel physical phenomenon which would have been difficult to detect otherwise. Perhaps the most noteworthy example of its utility is in highlighting  the role of anomalies in hydrodynamic behavior.

Indeed, following the work of \cite{Erdmenger:2008rm,Banerjee:2008th,Landsteiner:2011cp} which relied on holographic duality, it was understood that anomalies may play a prominent role in the dynamics of fluids \cite{Son:2009tf,Kharzeev:2010gr,Jensen:2012kj,Golkar:2012kb,Jensen:2013kka,Jensen:2013rga} (see also \cite{Kharzeev:2004ey}).  Since the latter analysis is broad and independent of holography, this paved the way for a variety of experimental attempts to observe anomalies in condensed matter systems, e.g., \cite{huang2015observation,zhang2016signatures,li2016negative,arnold2016negative,li2016chiral,guo2016large,gooth2017experimental,liang2018experimental,vu2021thermal} but see also \cite{li2016resistivity,luo2016anomalous,shen2016fermi} and \cite{baker1999linear,hu2005current}.

Of course, not all phenomena observed using the gauge gravity duality has a realistic counterpart which can be measured in the lab. More often than not, properties of holographic theories crucially rely on their highly supersymmetric structure or an a large $N$ (mean field) approximation. Indeed, the remarkable success of holographic methods in pointing out the role of anomalies in hydrodynamics which eventually lead to experimental data, was a result of a multi stage process. 

If the overall goal of a holographic computation is to set the stage for an experiment, then, unless the experiment mimics a supersymmetric, large $N$ theory, it should be the case that the holographic analysis captures a phenomenon which is robust and does not rely on the particular details of the underlying particle description of the theory. Thus, as a first step in relating holography to observations in the lab, one should check for robustness of the holographic result. By this we mean that the holographic result should not depend on the details of a particular dual theory. Rather it should be valid in a wide variety of holographic setups.

Once robustness within the holographic duality is made clear, one should check whether the holographic results may be observed outside the regime of the gauge gravity duality. Once this second step is fullfilled an experimental setup may be designed to validate the prediction which now goes beyond holographic correspondence.

In this work, we will demonstrate the robustness of holographic results pertaining to signatures of anomalies in out of equilibrium configurations, paving the way for a possible manifestation of such phenomenon in Weyl semi metals. Our starting point is the observation made in \cite{Ammon:2017ded,Abbasi:2018qzw,Bu:2016oba,Bu:2016vum,Bu:2018psl,Haack:2018ztx,Bu:2019mow,Meiring:2023wwi} that the quasi normal modes of magnetically charged black branes in asymptotically AdS${}_5$ geometries have long lived quasi normal modes at low temperatures as long as there is a strong enough dependence on the $U(1)$ anomaly. This observation was carried out in a very particular holographic setup which included gravity and a Maxwell field or fields. In \cite{Haack:2018ztx,Meiring:2023wwi} it was shown that the existence of a quasi normal mode can lead to an observable resonance of the anomalous $U(1)$ current. 

The goal of the current work is to bring the observation of \cite{Haack:2018ztx,Meiring:2023wwi} closer to the laboratory by demonstrating that the long lifetime of the appropriate quasi normal modes is holographically robust, in the sense described above, and will appear in generic holographic theories. 
In section \ref{S:Setup} we describe the holographic setup we use, its robustness and its limitations. Section \ref{sec:Analysis} contains the main results of this work. In it, we show under what conditions the low temperature quasi normal modes are long lived. We end with a new example demonstrating the validity of our results in section \ref{S:Example}. A discussion can be found in section \ref{S:Discussion}.

\section{Setup}
\label{S:Setup}
Consider an action involving a metric $g_{\mu\nu}$, a $U(1)$ field $A_{\mu}$ (with field strength $F=dA$), and other matter fields which we collectively denote by $\Phi$ whose only coupling to the $U(1)$ field is through a potential $V_F$,
\begin{equation}
\label{E:fullaction}
	S = S_{gravity+matter} - \int \sqrt{g} \left( \frac{1}{4}V_F(\Phi) F_{\mu\nu}F^{\mu\nu} + \lambda \epsilon^{\mu\nu\rho\sigma\tau} A_{\mu}F_{\nu\rho}F_{\sigma\tau} \right) d^5x\,.
\end{equation}
We assume that $S_{gravity+matter}$ supports asymptotically AdS solutions and that $\lambda$ is the coefficient of the five dimensional Chern-Simons term and is assumed not to appear in $S_{gravity+matter}$. Various extensions of this action will be discussed in section \ref{S:Discussion}.

We will be interested in the dynamics of \eqref{E:fullaction} in the limit where $\lambda$ is very large but $a_{\mu} = A_{\mu} \lambda$ is finite. In this case the dynamics of the $U(1)$ field decouple from the other fields so that the equations of motion are given by
\begin{align}
\begin{split}
\label{E:generalEOM}
	\frac{\delta}{\delta g^{\mu\nu}} S_{gravity+matter} & = 0\,, \\
	\frac{\delta}{\delta \Phi} S_{gravity+matter} & = 0\,, \\
	\nabla_{\mu} V_F f^{\mu\nu} &= 3 \epsilon^{\alpha\beta\gamma\delta\nu} f_{\alpha\beta}f_{\gamma\delta} \,,
\end{split}
\end{align}
where $f=da$. 

We are interested in a zero temperature, rotational and translation invariant solution (in the boundary directions) to the equations of motion \eqref{E:generalEOM}. We assume that these solutions exist and are locally stable. By local stability we mean that there are no zero modes associated with metric, matter field or gauge field fluctuations which break rotational or translational invariance at zero temperature. Note that we do not require global stability which implies that there could exist other, more stable, zero temperature solutions which are separated from the rotational and translation invariant one by a potential barrier. We provide an extended discussion on the implications of such solutions in section \ref{S:Discussion}. 

A zero temperature, rotation and translation invariant solution to the equations of motion can be brought to the form
\begin{equation}
\label{E:metricandphi}
	ds^2 = -f(r)dt^2 + 2 dr dt + L(r)^2 \sum_i (dx^i)^2\,,
	\qquad
	\Phi = \Phi(r)\,.
\end{equation}
The values of $f$, $L$ or $\Phi$ can not be determined without the explicit form of $S_{gravity+matter}$ which we have kept implicit in order to study the robustness of our result. However, since we are considering asymptotically AdS spacetimes, we require that, in our coordinate system 
\begin{equation}
	f(r) \xrightarrow[r\to\infty]{} r^2\,,
	\qquad
	L(r) \xrightarrow[r\to\infty]{} r\,,
\end{equation}
(where we have set the AdS radius at asymptotic infinity to unity)
and that $\Phi(r)$ must be finite in the $r\to\infty$ limit. Further, since this solutions presumably describe zero temperature configurations of gauge theories on $\mathbb{R}^{3,1}$ we expect that they posses some sort of horizon which we conveniently locate at $r=0$ by an appropriate shift of $r$. Thus, we expect that 
\begin{equation}
	f(r) \xrightarrow[r\to0]{} 0\,.
\end{equation}

The remaining equation for the gauge field takes the form
\begin{equation}
\label{E:fequation}
	\partial_{\mu} \left(L^3 V_F f^{\mu\nu}\right) = 3 \sqrt{g} \epsilon^{\alpha\beta\gamma\delta\nu} f_{\alpha\beta}f_{\gamma\delta}\,.
\end{equation}
Without the explicit value of $L$ and $V_F$ we can not solve \eqref{E:fequation} explicitly. As was the case for the metric, we keep $V_F$ implicit in order to study the dependence of the long lived quasi normal modes on the behavior of $V_F$. Nevertheless, regardless of the values of $V_F(\Phi(r))$ or $L(r)$, equation \eqref{E:fequation} supports a constant magnetic field solution, e.g., 
\begin{equation}
\label{E:gauge}
	a = B x^1 dx^2\,.
\end{equation}

Our main interest in this work is to study the quasi normal modes of large $\lambda$, low temperature, magnetically charged black branes. If the temperature is sufficiently low and $\lambda$ is sufficiently large (so that the metric doesn't backreact on the inhomogneous magnetic field) these black branes can be approximated by the zero temperature solutions described by \eqref{E:metricandphi} and \eqref{E:gauge}. With this in mind, we can approximate the quasi normal modes of these black branes by studying the modes associated with \eqref{E:metricandphi} and \eqref{E:gauge}. Of particular importance to this work are the quasi normal modes associated with fluctuations of $a_3$ which decouple from other modes due to rotational symmetry. Indeed, one can check that inserting $a = B x^1 dx^2 + \hbox{Re} (\delta a(r) e^{i \omega t})$ into \eqref{E:fequation} results in
\begin{equation}
\label{E:main}
	\delta a_3'' +\left(\frac{f'}{f} + \frac{L'}{L} + \frac{V_F'}{V_F} -\frac{2i\omega}{f} \right)  \delta a_3'+ \left(-\frac{576 B^2 }{f L^4 V_F^2} - i \frac{\omega}{f} \left( \frac{L'}{L} + \frac{V_F'}{V_F}\right)\right)  \delta a_3 = 0 \,.
\end{equation}
The boundary conditions for the quasi normal modes are that $\delta a_3$ vanishes at the asymptotic boundary, located at $r\to\infty$ and is finite at the zero temperature horizon located at $r=0$.

Equation \eqref{E:main} is the main result of this section. In the next section we will argue that, due to the Chern-Simons term, responsible for the $B^2$ dependence in \eqref{E:main}, there exist real solutions to \eqref{E:main} with real $\omega$. As discussed earlier, we expect that the low temperature quasi normal modes of the magnetically charged black hole asymptote to these real values. 

\section{Analysis}
\label{sec:Analysis}
We wish to solve \eqref{E:main} together with the boundary conditions that $\delta a_3$ vanishes at the asymptotic boundary and is finite at the horizon. Such solutions, if they exist, will be associated with a discrete set of values for $\omega$. To see this, we transform \eqref{E:main} into a Sturm Liouville type equation. Defining 
\begin{equation}
		\delta a_3 = e^{-i \int_r^{\infty} \frac{\omega dx}{f(x)}} k
\end{equation}
we find that \eqref{E:main} takes the form
\begin{equation}
\label{E:keq}
	\left( \left(f L V_F\right) k'\right)' - \frac{576 B^2}{L^3 V_F} k = - \omega^2 \frac{L V_F}{f} k\,.
\end{equation}

To understand the near boundary behavior of $\delta a_3$ we expand \eqref{E:keq} near the asymptotic boundary, located at $r\to\infty$. Since the metric is assumed to asymptote to the AdS metric, we have
\begin{subequations}
\label{E:boundaryasymptotics}
\begin{equation}
	f  = r^2 \left(1 + \mathcal{O}(r^{-1})\right)\,,
	\qquad
	L =  r \left(1 +\mathcal{O}(r^{-1})\right)\,.
\end{equation}
In addition we assume that
\begin{equation}
	V_F = V_{\infty} r^a\left(1+\mathcal{O}(r^{-1})\right)
\end{equation}
\end{subequations}
with $a$ a positive integer.
With this asymptotic behavior, the two linearly independent solutions to \eqref{E:keq} behave as
\begin{equation}
\label{E:NBasymptotics}
	k = 1+\ldots\,,
	\qquad
	\hbox{or}
	\qquad
	k = r^{-2-a}+\ldots \,,
\end{equation}
near the asymptotic boundary located at $r\to\infty$ and for $a>-2$. (We will see shortly that in order to ensure that $\omega^2$ be real, we should set $a>0$). In order to satisfy the boundary condition $\delta a_3(\infty)=0$ we choose the solution for which $k(\infty)=0$.

Near the horizon, located at $r=0$, we assume that
\begin{equation}
\label{E:asymptotics}
	f = f_0 r^{1+z}\left(1+\ldots\right) \,
	\qquad
	L = L_0 r^y\left(1+\ldots\right)\,
	\qquad
	V_F = V_0 r^x\left(1+\ldots\right)\,,
\end{equation}
with $x>0$, $y>0$ and $z > 0$. Some examples of geometries of this type include
\begin{enumerate}
	\item Non extremal AdS black brane with 
	\begin{equation}
		z=0\,, \qquad y=0\,, \qquad  x=0
	\end{equation}
	\item Extremal, Reissner-Nordstrom AdS black brane
	\begin{equation}
		z=1\,, \qquad y=0\,, \qquad x=0
	\end{equation}
	\item AdS space with a Poincare horizon
	\begin{equation}
		z=1\,, \qquad y=1\,, \qquad x=0
	\end{equation}
	\item Lifshitz geometries
	\begin{equation}
		z = 2\zeta - 1\,, \qquad y=1,\, \qquad x=0\,.
	\end{equation}
\end{enumerate}
In any case, to understand the behavior of $k$ near the horizon we consider
\begin{equation}
	k'' + \frac{1+x+y+z}{r}k' - \frac{576 B^2}{f_0 L_0^4 V_0^2\,r^{1+2x+4y+z}} k = - \frac{\omega^2}{f_0^2\,r^{2+2z}} k\,.
\end{equation}

To understand the near horizon behavior of $k$ we separate into cases.
If $2x+4y>1+z$ and $B \neq 0$, the $B^2$ term dominates over the $\omega^2$ term in the small $r$ limit. Then $2x+4y+z -1 >2z >0$ and the asymptotic behavior of $k$ for small $r$ takes the form
\begin{equation}
\label{E:kasymptotics}
	k \sim 
		r^{-\frac{1}{4}(z+1-2y)} \exp\left(\pm \frac{48 B}{\sqrt{f_0}L_0^2 V_0(2x+4y+z-1)} r^{-\frac{1}{2} \left(2x+4y+z-1\right)} \right) 
\end{equation}
implying that solutions which are regular at the horizon satisfy $k(0)=0$.
If $2x+4y < 1+z$ or if $B=0$ the asymptotic behavior of $k$ near $r=0$ takes the form
\begin{equation}
\label{E:kasymptotics2} 
	k \sim r^{-\frac{1}{2}(x+y)} \exp\left(\pm i \sqrt{\frac{\omega^2}{f_0 z}}r^{-z}\right)\,.
\end{equation}
If $2x+4y = 1+z$ we get
\begin{equation}
\label{E:kasymptotics3}
	k \sim r^{-\frac{1}{2}(x+y)} \exp \left( \pm i\sqrt{\frac{\omega^2}{f_0^2 z^2} - \left(\frac{48 B}{\sqrt{f_0}L_0^2 V_0 2 z}\right)^2}  r^{-z} \right)\,.
\end{equation}
The choice of sign in \eqref{E:kasymptotics2} or \eqref{E:kasymptotics3} depends on whether we require incoming or outgoing boundary conditions at the horizon.

When $2x+4y > 1+z$ the problem of finding an $\omega$ which solves \eqref{E:keq} reduces to a (singular) Sturm-Liouville problem. Indeed, \eqref{E:keq} and its associated boundary conditions become
\begin{subequations}
\label{E:SLproblem}
\begin{equation}
\label{E:SLequation}
	\left(p k' \right)' + q k = -\Omega m k
\end{equation}
where
\begin{equation}
\notag
	p = f L V_F,\qquad
	q = -\frac{576 B^2}{L^3 V_F},\qquad
	m = \frac{L V_F}{f}\,,\qquad
	\Omega= \omega^2
\end{equation}
with the boundary conditions
\begin{equation}
\label{E:SLbcs}
	k(0)=0\,, \qquad k(\infty)=0\,.
\end{equation}
\end{subequations}

If $k$ were defined on a finite interval on which $p>0$ and $m>0$ we would have been guaranteed that there exist a sequence of real $\Omega_n$'s, $\Omega_1 < \Omega_2 < \ldots$ for which a solution to \eqref{E:SLproblem} with $\Omega=\Omega_n$ exists. Since $k$ is defined on a semi infinite interval and since $p$ and $m$ are not strictly positive at $r=0$ and may diverge at $r=\infty$, then \eqref{E:SLproblem} is classified as a singular Sturm-Liouville equation. As we will now show the spectrum is, nevertheless, bounded from below and positive.

First, we argue that if a solution to \eqref{E:SLproblem} exists then $\Omega$ must be non negative.
To see this we multiply the Sturm Liouville equation \eqref{E:SLequation} by $k^{\star}$ obtaining
\begin{align}
\begin{split}
\label{E:omegaval}
	\Omega \int_0^{\infty} |k|^2 m dr &= -\int_0^{\infty} \left( (p k')' + q k \right)k^* dr \\
		& = \int_0^{\infty} \left( p |k'|^2 - q |k|^2 \right) dr - p k' k^* \big|_0^{\infty}\,.
\end{split}
\end{align}
Note that both integrals are well defined: near the horizon, at $r=0$, $m$, $p$ and $q$ diverge at worst as an inverse power of $r$ whereas $k$ approaches zero exponentially fast. Likewise, near the asymptotic boundary at $r\to\infty$, $m \sim r^{-1+a}$, $p\sim r^{3+a}$ and $q \sim r^{-3-a}$ whereas $k \sim r^{-2-a}$. Now, given this asymptotic behavior, the last term on the right hand side of \eqref{E:omegaval} vanishes. Since $p$ and $m$ are positive in $r\in(0,\infty)$ and $q$ is negative on it, we find
\begin{equation}
	\Omega \geq 0\,,
\end{equation}
If $\Omega=0$ then there exists a zero mode implying that the zero temperature solution is unstable to a rotational symmetry breaking mode ($a_3$). Assuming rotational symmetry is locally stable we find that if a solution to \eqref{E:SLproblem} exists then $\Omega$ must be positive.

It remains to show that solutions to \eqref{E:SLproblem} exist. Note that the asymptotic behavior of $k$ near the origin $r=0$ given by \eqref{E:kasymptotics} specifies the solution all the way to the AdS boundary located at $r \to \infty$ up to an overall multiplicative constant. Let us track one such a solution up to some radial parameter $\epsilon \ll 1$. We tune $\epsilon$ so that both $k(\epsilon)$ and $k'(\epsilon)$ are non zero. We denote $k(\epsilon)=K$ and $k(\epsilon)/k'(\epsilon)=\theta$. Note that $\theta$ is determined by the requirement that the solution vanishes at the origin and that $K$ can be tuned by an appropriate choice of the overall multiplicative constant of which $\theta$ is independent.

We can now solve the Sturm Liouville problem associated with \eqref{E:SLequation} on the interval $r\in[\epsilon,L]$ with $L \gg 1$ with boundary conditions
\begin{subequations}
\label{E:bcs2}
\begin{align}
\label{E:SLbceps}
	\frac{k(\epsilon)}{k'(\epsilon) } &=  \theta \,, \\
\label{E:SLbcL}
	k(L) &= k_L\, 
\end{align}
\end{subequations}
for some non zero value of $k_L$. Since this is a standard, non singular, Sturm Liouville problem, we are guaranteed that there exist a discrete set of eigenvalues, $\Omega_0 < \Omega_1 < \ldots$ which solve \eqref{E:SLequation} with \eqref{E:bcs2}. Further, we are guaranteed that if we integrate such a solution all the way to $r=0$ we will obtain $k(0)=0$. To see this, note that we may always tune the free multiplicative factor discussed in the previous paragraph, characterized by $K$, so that gluing the solution defined in the region $0 \leq r \leq \epsilon$ with the solution to \eqref{E:SLequation} with \eqref{E:bcs2} generates a continuous and differentiable function on $0 \leq r \leq L$.

The solution obtained by solving \eqref{E:SLequation} with the boundary conditions \eqref{E:bcs2} can be integrated all the way from $L$ to $\infty$. If $L$ is sufficiently large, then from the near boundary asymptotic behavior of $k$ given in  \eqref{E:NBasymptotics}, the behavior of the solution in this region may be approximated by
\begin{equation}
	k(r) = k_L + \mathcal{O}\left(L^{-\min\{2,2+a\}}\right)\,.
\end{equation}
Thus, the eigenvalues $\Omega_n$ solve \eqref{E:SLequation} with boundary conditions given by \eqref{E:SLbceps} and 
\begin{equation}
\label{E:SLbcinf1}
	k(\infty) = k_L+ \mathcal{O}\left(L^{-\min\{2,2+a\}}\right)\,.
\end{equation}
Now, we could use the same argument to show that there is a different set of eigenvalues $\tilde{\Omega}_n$ which solve \eqref{E:SLequation} but with boundary conditions \eqref{E:SLbceps} and
\begin{equation}
	k(\infty) = -k_L+ \mathcal{O}\left(L^{-\min\{2,2+a\}}\right)\,.
\end{equation}
From continuity (in $k_L$) we should also have a set of eigenvalues which solve \eqref{E:SLequation} with
\begin{equation}
\label{E:SLbcs3}
	\frac{k(\epsilon)}{k'(\epsilon) }= \theta\,,
	\qquad
	k(\infty) = 0\,.
\end{equation}
As before, we are guaranteed that if we integrate such a solution to the origin we will find that $k(0)=0$.

Thus, we have shown that solutions to \eqref{E:SLproblem} exist. Our previous argument guarantees that there is a minimal positive value of $\Omega$. Positivity of $\Omega$ implies that the associated frequencies, $\omega$, are real.

This is the main result of this work. For large values of $\lambda$, as long as the asymptotic behavior of the zero temperature configuration, as specified by \eqref{E:asymptotics}, satisfies $2x+4y>1+z$ then the quasi normal modes of the gauge field will have a vanishing imaginary component. If the zero temperature configuration is locally stable then the real component of the modes will be non trivial implying long lived oscillations of the associated dual current \cite{Haack:2018ztx,Meiring:2023wwi}.

If $2x+4y \leq 1+z$ then the asymptotic behavior of $k$ will depend on the frequency $\omega$ (see \eqref{E:kasymptotics2} and \eqref{E:kasymptotics3}. The associated boundary value problem will no longer be a Sturm Liouville type problem and it is more difficult to analyze the resulting spectrum.

\section{An explicit low temperature realization}
\label{S:Example}

In the previous section we have obtained the condition required in order that the zero temperature quasi normal modes of the gauge field become real as the Chern Simons coupling $\lambda$ becomes large. Writing the near horizon asymptotics for the metric as in \eqref{E:asymptotics}, we need that 
\begin{equation}
\label{E:condition}
	2x+4y>1+z
\end{equation}
in order for the quasi normal modes to become real. Note that the entropy of the black hole which asymptotes to the geometry in \eqref{E:asymptotics} is finite only when $y\geq 0$ and vanishes for $y>0$. Thus, in the absence of a scalar potential which vanishes sufficiently fast, non vanishing entropy at zero temperature will prevent the appearance of long lived quasi normal modes. In \cite{Haack:2018ztx,Meiring:2023wwi} a particular realization of long lived quasi normal modes were studied where the zero temperature geometry was empty AdS. In what follows we discuss a more intricate setup where the zero temperature configuration is a domain wall solution interpolating between two AdS geometries.

Consider the following action involving gravity and a real scalar field $\phi$ (see \eqref{E:fullaction})
\begin{equation}
\label{E:saction}
	S_{gravity+matter} = \frac{1}{16\pi G_N} \int \sqrt{-g} \left(R + 12 - \frac{1}{2} (\nabla \phi)^2 - V(\phi)\right)d^5x
\end{equation}
with
\begin{equation}
\label{E:Vval}
	V(\phi) = -\frac{1}{2} m^2 \phi^2 + \frac{u^2}{4} \phi^4\,.
\end{equation}
The homogenous and isotropic (in the boundary directions) solution for the metric and scalar field are of the form
\begin{align}
\begin{split}
\label{E:ansatz}
	ds^2 &= -A(r)e^{-\chi(r)}  dt^2 + \frac{1}{A(r)} dr^2 + r^2 dx_i^2 \\
	\phi & = \phi(r)\,.
\end{split}
\end{align}
Inserting \eqref{E:ansatz} into the equations of motion derived from \eqref{E:saction} we find
\begin{align}
\begin{split}
\label{E:sEOM}
	\phi'' + \left(\frac{3}{r} + \frac{A'}{A} - \frac{1}{2} \chi'\right)\phi' - \frac{V'}{A} & = 0 \\
	\left(r^5 e^{\frac{\chi}{2}} \left(\frac{A e^{-\chi}}{2} \right)'\right)' & = 0 \\
	\chi' + \frac{1}{3} r \phi' & = 0\,.
\end{split}
\end{align}

Assuming a black hole solution with an event horizon at $r=r_h$ such that $A(r_h)=0$, we can integrate the second equation in \eqref{E:sEOM} and evaluate it at $r = r_h$ leading to
\begin{equation}
\label{E:Qval}
	r^5 e^{\frac{\chi}{2}} \left(\frac{A e^{-\chi}}{r^2} \right)' = 16 \pi s T 
\end{equation}	
where $s = \frac{1}{4}r_h^3$ is the entropy per unit volume and $T= \frac{A'e^{-\frac{\chi}{2}}}{4\pi}$ is the Hawking temperature. We are interested in the zero temperature solution for which \eqref{E:Qval} reduces to 
\begin{equation}
\label{E:T0sol}
	\frac{A e^{-\chi}}{r^2} = 1\,.
\end{equation}
The right hand side of \eqref{E:T0sol} was chosen to be one by scaling the time coordinate appropriately, viz. the line element takes the form,
\begin{equation}
\label{E:zeroTLE}
	ds^2 = -r^2 dt^2 + \frac{dr^2}{A} + r^2 \sum_i dx_i^2\,.
\end{equation}
The equations of motion reduce to
\begin{subequations}
\label{E:zeroT}
\begin{equation}
\label{E:zerophi}
	\phi'' + \frac{5\phi'}{r} - \frac{1}{6} (\phi')^3 - \frac{(r^2 (\phi')^2-24)V'}{r^2(2 V-24)}  = 0 
\end{equation}
and the metric components are determined algebraically from the scalar field
\begin{align}
\label{E:Azero}
	A & =  \frac{r^2(2 V-24)}{r^2 (\phi')^2-24}\\
	\chi & = - \ln\frac{r^2}{A}\,.
\end{align}	
\end{subequations}

As one may expect, the zero temperature solutions to \eqref{E:zeroT} are domain walls interpolating between the $\phi=0$ unstable solution at $r\to\infty$ and one of the $|\phi| = \frac{m}{u}$ solutions at $r=0$. We show this explicitly in appendix \ref{A:ZeroTscalar}. Thus, near the horizon, at $r=0$, we have
\begin{equation}
	A = \left(1-\frac{V(m/u)}{12}\right) r^2 + \ldots \,.
\end{equation}
We can now go to Eddington-Finkelstein coordinates where the zero temperature line element given in \eqref{E:zeroTLE} takes the form
\begin{equation}
\label{E:zeroTEF}
	ds^2 = -\alpha^2(\rho) dv^2 + 2 dv d\rho + \alpha^2(\rho) \sum_i dx_i^2 
\end{equation}
where
\begin{equation}
\label{E:alphaval}
	\alpha = \frac{\sqrt{12-V(m/u)}}{2\sqrt{3}}\rho\,.
\end{equation}
Comparing \eqref{E:zeroTEF} to the line element in \eqref{E:metricandphi} we can parameterize the asymptotic behavior of the metric using \eqref{E:asymptotics} so that the condition for real quasi normal modes \eqref{E:condition} becomes
\begin{equation}
\label{E:VFconstraint}
	x > -1
\end{equation}
where $x$ is the behavior of a possible scalar potential coupling to the gauge field, c.f., \eqref{E:fullaction}, characterized by $V_F$. Assuming that $V_F$ is finite around the minimum of the potential $V$, condition \eqref{E:VFconstraint} will be satisfied. We demonstrate this behavior numerically, in a slightly simplified setting in appendix \ref{A:lowTQNM}.

\section{Discussion}
\label{S:Discussion}

The low temperature quasi normal modes of a large class of asymptotically AdS black branes associated to (generic) holographic theories with an anomalous $U(1)$ current whose strength is sufficiently large will have vanishingly small imaginary components. At least as long as the qualifiers we have mentioned in the main text are met. These qualifiers are that the zero temperature solution is locally stable to spontaneous rotation symmetry breaking and that, roughly, the entropy of the zero temperature solution, or more precisely, that the asymptotic near horizon geometry is such that the inequality \eqref{E:condition} is satisfied.

Let us discuss these criteria in some detail. It is known that the presence of a Chern-Simons term in the bulk may trigger spontaneous symmetry breaking. See, for instance, \cite{Ooguri:2010kt,Nakamura:2009tf,Donos:2011bh,Bergman:2011rf,Donos:2012wi,Ovdat:2014ipa,Withers:2014sja}. In all these instances spontaneous symmetry breaking occurs at some critical temperature $T_c$. Our criterion for real quasi normal modes is that there is no such transition at zero temperature, i.e., $T_c \neq 0$. As far as we know, there are no examples of such transitions in holography. The cautious reader might, however, be worried about the possible implications of our result if rotational invariance is broken at some non zero $T_c$. Indeed, if there is a phase transition to a non homogenous phase at $T_c>0$ then our analysis will no longer be relevant for temperatures below $T_c$. However, if $T_c$ is sufficiently small then we should be able to observe quasi normal modes with increasingly small imaginary components even for small $T>T_c$. Since an estimate of the asymptotic behavior of the quasi normal modes and the dependence of their imaginary component on the temperature is model dependent, we can not say much more about the behavior of the system when $T_c>0$. Such configurations need to be studied on a case by case basis.

Zero temperature black brane solutions with non zero entropy are abundant. The Reissner-Nordstrom black brane solution is an example of such. In the event where the black hole entropy is non zero at zero temperature, or, more precisely when \eqref{E:condition} is not met, then the low temperature quasi normal modes are not expected to have a vanishingly small imaginary components. It is interesting to contrast this with the recent findings of \cite{Iliesiu:2020qvm} which shows that extremal non supersymmetric black holes have zero entropy due to quantum effects. It is worth exploring whether the quasi normal modes will be affected by this behavior. Regardless, we expect that when making the transition to realistic, non holographic systems, the third law of thermodynamics will hold.

As discussed in the introduction, the vanishingly small imaginary component of low temperature quasi normal modes is not new \cite{Ammon:2017ded,Abbasi:2018qzw,Bu:2016oba,Bu:2016vum,Bu:2018psl,Haack:2018ztx,Bu:2019mow,Meiring:2023wwi}. What we have demonstrated in this work is that this phenomenon is robust suggesting that it may be realized outside the realm of holography. If this is the case, materials whose effective description incorporates an anomalous $U(1)$ current may posses long lived quasi normal modes at sufficiently low temperatures. Such modes may be observed by, say, exciting the material via a quench whose Fourier modes are compatible with the quasi normal modes of the system. Exciting such a long lived mode implies that a long lived current will be generated whose lifetime is much longer than the lifetime of the quench. We refer the reader to \cite{Haack:2018ztx,Meiring:2023wwi} for an extended discussion.

To excite the lowest lying quasi normal mode, $\omega_0$, we need to provide an estimate of its value. This is difficult in a generic setting. Its dependence on the various dimensionful parameters of the theory are probably model dependent. Nevertheless we can estimate $\omega_0$ using dimensional analysis. Denoting the anomaly coefficient by $\lambda$, the magnetic field by $B$, and the other dimensional parameters of the theory by $\vec{\Lambda}$ we expect that $\omega_0 = \omega_0(\lambda\,,B\,, \vec{\Lambda})$. Recall that we have assumed that $\lambda$ is sufficiently large that the dynamics of the gauge field decouple from gravity and the other matter content of the theory, c.f., \eqref{E:generalEOM}. In fact, we have implicitly encoded the anomaly coefficient in the magnetic field by taking the limit where $\lambda$ is large and the gauge field is small. Thus, we expect that $\omega_0 = \omega_0(\lambda\,B\,,\vec{\Lambda})$. While it is difficult to estimate the functional dependence of $\omega_0$ on $\lambda B$ and $\vec{\Lambda}$ we do note that $B$, the magnetic field, is a tuneable parameter whereas $\vec{\Lambda}$ is a parameter that describes the dynamics of the fixed zero temperature solution. If we tune $B$ to be sufficiently large so that $B \ll |\vec{\Lambda}|$ in appropriate units then it must be the case that
\begin{equation}
	\omega_0 = C \frac{B \lambda c^{\frac{3}{2}}}{\hbar^{\frac{1}{2}}}
\end{equation}
where $C$ is a numerical constant, $c$ is the effective speed of light, and $\hbar$ is Planck's constant.

In an effective condensed matter system we may set $c$ to the Fermi velocity, roughly of order $10^6$ m/s \cite{dolui2015theoryweylorbitalsemimetals} so that
\begin{equation}
	\frac{\omega_0}{1 \, \hbox{THz}} \sim \frac{B N}{1 \, \hbox{Gauss}}  
\end{equation}
where we have assumed that $C$ is of order $2$ and have set $\lambda = \frac{N}{8\pi^2}$, $N$ denoting the number of species of anomalous fermions in the effective theory. This estimate is in the same ballpark as that of \cite{Haack:2018ztx} which included finite temperature effects in a particular holographic setup. There is an exciting possibility that magnetic fields and frequencies of these orders of magnitude may be generated in a controlled environment.

There are several interesting directions that one may further pursue. First, one can extend the class of theories for which long lived quasi normal modes are present. In this work we have considered bulk actions of the form given in \eqref{E:fullaction} where the matter content is unspecified but also uncharged under the single anomalous $U(1)$ field represented by the gauge field $A_{\mu}$. For instance, one can discuss configurations with several anomalous $U(1)$ fields, viz.,
\begin{equation}
\label{E:generalization1}
	S = S_{gravity+matter} - \int \sqrt{g} \frac{1}{4} V_{F\,ij} F^i_{\mu\nu}F^{j\mu\nu} + \lambda \epsilon^{\mu\nu\rho\sigma\tau}_{ijk}A_{\mu}^i F_{\nu\rho}^j F_{\sigma\tau}^k\,.
\end{equation}
where $i$ runs from $1$ to $N_F$ (see, for instance, \cite{Donos:2011qt,Meiring:2023wwi,Wang:2024bli,Deshpande:2024itz}). In general, one can still apply a large $\lambda$ limit to decouple the equations of motion for the gauge fields (which are presumably anomalous) from the matter content of the theory. The resulting equations of motion would be a modified version of \eqref{E:fequation}. If the $V_{ij}$ are independent on $\Phi$ then the resulting quasi normal modes have been shown to approach the real axis, similar to what we have found in this work \cite{Meiring:2023wwi} (see also appendix \ref{A:lowTQNM} for an extension of these results). It would be interesting to extend that discussion to general $V_{ij}$.

Another extension of our result is to consider matter which is charged under the ('t Hooft) anomalous $U(1)$ symmetry. If the matter has charge $e$ then we may scale $e$ to infinity while keeping $q = e/\lambda$ finite. This reproduces \eqref{E:generalEOM} if the currents $J^{\mu} = \frac{\delta}{\delta A_{\mu}} S_{gravity+matter}$ vanish. Otherwise, it seems that a more involved analysis is required.

Another interesting extension of our result would be to check the dependence of the imaginary component of the quasi normal modes on temperature and on the coupling $\lambda$. In \cite{Meiring:2023wwi} perturbation theory in $\lambda$ and in temperature was used to compute subleading corrections to the (real) zero temperature quasi normal modes. A similar computation can be carried out in this instance which might be helpful in identifying the temperature at which the imaginary components of the quasi normal modes are sufficiently small.

Perhaps the most interesting extension of these results would be to identify a field theoretic understanding, independent of the gauge gravity duality, for the existence of long lived modes or quasi particles, which the temperature is sufficiently low and the strength of the anomaly is sufficiently large. We hope to report on such findings in the near future.

\section*{Acknowledgements}
This work is supported in part by an ISF-NSFC grant 3191/23, a BSF grant 2022110 and an ISF grant 2916/24, and grant No. 1417/21, by the German Research Foundation through a German-Israeli Project Cooperation (DIP) grant “Holography and the Swampland”, by Carole and Marcus Weinstein through the BGU Presidential Faculty Recruitment Fund, by the ISF Center of Excellence for theoretical high energy physics and by the ERC Starting Grant dSHologQI (project number 101117338).

\begin{appendix}

\section{Zero-temperature hairy black holes}
\label{A:ZeroTscalar}
In section \ref{S:Example} we considered the Einstein-scalar system given by the action \eqref{E:saction} whose zero temperature solution takes the form \eqref{E:zeroTLE} as long as \eqref{E:zeroT} are satisfied. In the main text we have argued that as long as the (zero temperature) horizon is located at $r=0$ then the condition for the existence of long lived quasi normal modes, \eqref{E:condition}, is satisfied (assuming $V_F$ is non divergent at the horizon). In this section we show that the horizon must be located at $r=0$ and not at some non zero value of the radial coordinate. (Note that if the horizon was located at a finite value of the radial coordinate then the entropy would be non zero when the temperature is zero in which case \eqref{E:condition} would not be satisfied implying that long lived quasi normal modes do not neccessarily exist.)

Recall the zero temperature line element is given by \eqref{E:zeroTLE} and the equations of motion for the scalar field and metric are given by \eqref{E:zeroT}. The essential features that we will need of the potential $V$ given in \eqref{E:Vval} is that it is $W$ shaped. More precisely, that it has a local maximum at $\phi=0$, $V(0)=0$, and a pair of minima after which $V(\phi)$ increases monotonically.

We will insist on boundary conditions such that the scalar field vanishes near the boundary located at large $r$, $r\to\infty$. Taylor expanding the scalar field near the boundary we find
\begin{equation}
\label{E:phibexpansion}
	\phi = \Lambda r^{-(4-\Delta)}\left(1+ \ldots\right) + \langle O \rangle r^{-\Delta}(1+\ldots)
\end{equation}
where $\ldots$ involve higher powers of $r$ and $\ln r$ and $\Delta = 2+\sqrt{4-m^2}$ and $m$ is the coefficient of the quadratic term in the potential, $V = -\frac{1}{2}m^2\phi^2+\ldots$. The parameters $\Lambda$ and $\langle O \rangle$ are integration constants. Under the gauge gravity duality  $\langle O \rangle$ is proportional to the expectation value of the operator, $O$, dual to $\phi$ and $\Lambda$ to a source term for $O$. For some values of $\Delta$ the role of $\langle O \rangle$ and $\Lambda$ may be flipped \cite{Klebanov:1999tb}. 

Let us suppose that the outermost event horizon is located at some $r=r_h>0$. Then, \eqref{E:Qval} implies that
\begin{equation}
	\frac{A'(r_h)}{\sqrt{A(r_h)}}r_h = 0
\end{equation}
suggesting that
\begin{equation}
	A(r) \sim (r-r_h)^{2+\epsilon}
\end{equation}
near $r = r_h$ with $\epsilon>0$. It then follows from the equation of motion for $A$ that
\begin{equation}
	\phi'(r) \sim (r-r_h)^{-\frac{1}{2}}\,.
\end{equation}
Inserting this into the equation of motion for $\phi$ and evaluating at $r=r_h$ we find that
\begin{equation}
\label{E:Vmin}
	V'(\phi(r_h))=0
\end{equation}
so that the value of the field at the horizon is at an extremum of the potential.

From the asymptotic behavior near the horizon and near the boundary we have
\begin{align}
\begin{split}
	r^2 (\phi^{\prime})^2 \xrightarrow[r\to\infty]{} & 0\,, \\
	r^2 (\phi^{\prime})^2 \xrightarrow[r\to r_h]{} & \infty\,.
\end{split}
\end{align}
Thus, there should exist an $r_0$ such that
\begin{equation}
\label{E:defr0}
	r_0^2 (\phi'(r_0))^2 = 24\,.
\end{equation}
But continuity of $A$ demands that  for such an $r_0$ we have
\begin{equation}
\label{E:V12}
	V(\phi(r_0))=12\,.
\end{equation}
For convenience we denote $\phi(r_0)=\phi_{12}$. 
To summarize, so far we have seen that if $r_h > 0$ then the scalar field will overshoot the minimum of the potential and will reach $\phi(r_0) = \phi_{12}$ for $r_h<r_0<\infty$. 

Since $\phi'(r_0) = \sqrt{24}/r_0\neq 0$ the scalar field will necessarily overshoot $\phi_{12}$. Once $\phi>\phi_{12}$ then $V>12$ and since $A$ is continuous and can not vanish unless $r=r_h$, then neccessarily $|\phi'(r)| > \sqrt{24}/r$ for all $r_h<r<r_0$. But then, there is no $r<r_0$ for which $\phi'(r)=0$ so the scalar field can not bounce back to the minimum of the potential contradicting \eqref{E:Vmin}. Hence, $r_h=0$.

In addition to the somewhat abstract argument described above we have also constructed numerical solutions to \eqref{E:zeroT} where the scalar field interpolates between the maximum of the potential near the AdS boundary to its minimum at $r=0$. 

In order to obtain a numerical solution to \eqref{E:zeroT} we first solve \eqref{E:zerophi} and then use the solution to obtain $A$ algebraically via \eqref{E:Azero}. The boundary conditions we should impose on the equation of motion for $\phi$ can be determined by studying the asymptotic behavior of $\phi$ near the boundary located at $r\to\infty$ and near the horion $r=0$. 

Near the boundary the asymptotic behavior of the potential is given by \eqref{E:phibexpansion}. Near the horizon, located at the minimum of the potential at $r=0$, the scalar field takes the asymptotic form
\begin{equation}
\label{E:phih}
	\phi = \frac{m}{u} + c_1 r^{-\Delta_{IR}}(1 + \ldots) + c_2 r^{-(4-\Delta_{IR})}(1 + \ldots)
\end{equation}
where $\Delta_{IR} = 2 + \sqrt{4+\frac{96 m^2 u^2}{m^4+48u^2}}$. 

To obtain solutions to the scalar equation in \eqref{E:zeroT} which are finite at the horizon, we need to set $c_1=0$ in \eqref{E:phih}. This leaves us with a one parameter family of solutions which is compatible with invariance of \eqref{E:zerophi} under a rescaling of the radial coordinate 
\begin{equation}
\label{E:rescaling}
	r\to r/r_*\,.
\end{equation}
Put differently, given a solution to \eqref{E:zerophi} with asymptotic behavior $c_2=1$ we may generate a solution with $c_2=c$ by rescaling the radial coordinate as in \eqref{E:rescaling} with $c = r_*^{(4-\Delta_{IR})}$. The invariance \eqref{E:rescaling} is a reflection of the underlying conformal symmetry before its explicit breaking by the source term $\Lambda$ in \eqref{E:phibexpansion}. All dimensionful CFT data are measured relative to the scale $\Lambda$. At the end of the day, a zero temperature solution to \eqref{E:zeroT} can be obtained by integrating \eqref{E:zerophi} from the horizon located at $r=0$ with boundary data given by \eqref{E:phih} with $c_1=0$ and (say) $c_2=1$.

\begin{figure}
\centering
\includegraphics[width=1\linewidth]{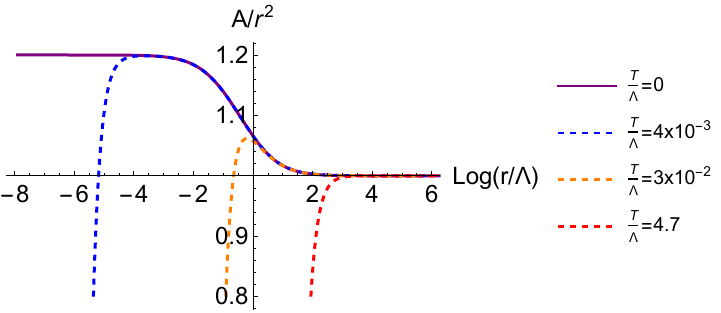}
\caption{A plot of the zero temperature domain wall geometry given by the line element \eqref{E:zeroTLE} whose dynamics are determined by the action \eqref{E:saction} and scalar potential \eqref{E:Vval} with $m^2=3$ and $u=\sqrt{15}/4$ (purple). Dashed lines demonstrate that finite temperature solutions approach the zero temperature one as the temperature is lowered. Here $T$ is the temperature and $\Lambda$ is the source term for the scalar which can be read off the asymptotic behavior of $\phi$ via \eqref{E:phibexpansion}.}
\label{F:zero}
\end{figure}

In figure \ref{F:zero} we have plotted the zero temperature solution to the metric component $A$ determined through \eqref{E:Azero} after having solved \eqref{E:zerophi}. We have checked that finite temperature solutions asymptote to this zero temperature solution as the temperature is lowered. Finite temperature solutions have been obtained by solving \eqref{E:sEOM}. These equations can be reduced to a coupled set differential equations where a second order equation for $\phi$ is coupled to a first order equation for $A$. The required asymptotic behavior for $A$ near the boundary, $A/r^2 \xrightarrow[r\to\infty]{}1$, together with a scaling symmetry similar to \eqref{E:rescaling} implies a one parameter family of solutions which could be parameterized by the temperature, or, the value of $\phi$ at the horizon.

\section{Numerical analysis of QNM's at non zero $T$.}
\label{A:lowTQNM}

We have discussed the conditions for which the low temperature quasi normal modes of magnetically charged black branes will have a small imaginary part. This is an extension of the work carried out in \cite{Haack:2018ztx,Meiring:2023wwi} where the rate at which the quasi normal modes approach their zero temperature limit was also studied. In this section we carry out a similar analysis.

Magnetic black branes in AdS${}_5$ are non isotropic which makes numerical solutions more involved due to the additional metric components which need to be tracked. As mentioned in section \ref{S:Discussion}, one way to bypass this minor difficulty is to consider multiply charged black branes.

Therefore, in this section we consider a modification of \eqref{E:fullaction} which takes the form
\begin{equation}
\label{E:Sexample}
	S = S_{gravity+matter} + \int \sqrt{-g} \left(  - \frac{1}{4} \sum_{i=1}^3\delta_{ij} F^i{}_{\mu\nu}F^{j\,\mu\nu} + \lambda \epsilon^{\mu\nu\rho\sigma\tau} A^1{}_{\mu}F^2_{\nu\rho}F^3_{\sigma\tau} \right) d^5x
\end{equation}
where $S_{gravity+matter}$ takes the form \eqref{E:saction} with \eqref{E:Vval} and $m^2=3$ and $u=1$. We are interested in studying the behavior of the quasi normal modes associated with magnetically charged configurations at large $\lambda$. The ansatz we use for the metric, gauge fields and scalar are 
\begin{align}
\begin{split}
\label{E:Aexample}
	ds^2 &= -A(r) e^{-\chi{r}}dt^2 + \frac{dr^2}{A(r)} + r^2 \sum_i (dx^i)^2\,,\\
	\phi &=\phi(r) \,,\\
	A_{\mu}^1 dx^{\mu} &= B y dz\,,
	A_{\mu}^2 dx^{\mu} =B z dx\,,
	A_{\mu}^3 dx^{\mu} = B x dy\,.
\end{split}
\end{align}
The magnetic field has been chosen so that the solution respects an $SO(3)$ symmetry which slightly simplifies our analysis.

\begin{figure}
\centering
\includegraphics[width=0.45\linewidth]{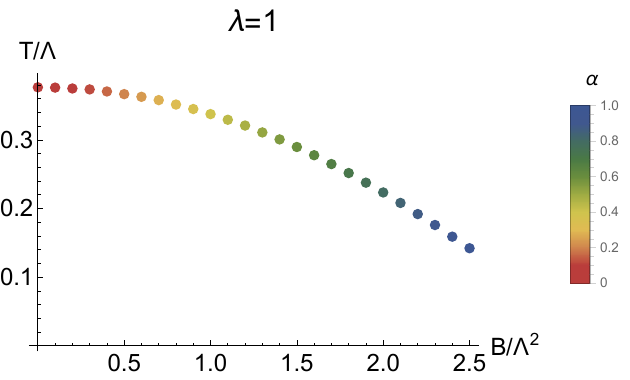}
\hfill
\includegraphics[width=0.45\linewidth]{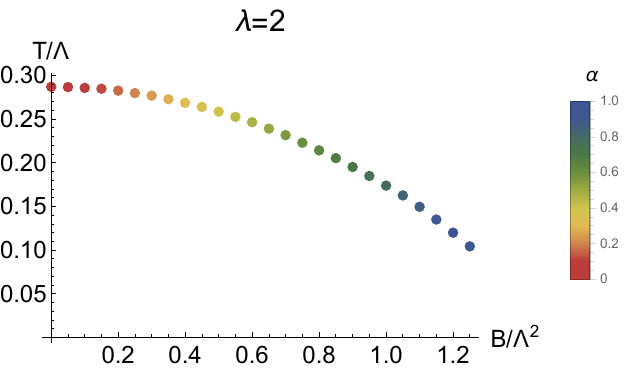}
\caption{Plots of the parametric values of the temperature and magentic field for the solutions we have found to the equations of motion associated with a magnetically charged black brane following from varying \eqref{E:Sexample} and using the ansatz \eqref{E:Aexample}. The solutions lie on curves on the $T/\Lambda$,$B/\Lambda^2$ plane which we parameterize by $\alpha$. (See the legend.)}
\label{F:BTcurve}
\end{figure}

We will not write out the equations of motion which follow from varying \eqref{E:Sexample} but merely mention that they are similar to \eqref{E:sEOM} with extra factors of $B$ associated with the presence of the magnetic field. We have solved these equations of motion by integrating outward from the horizon, very similar to the construction in \ref{A:ZeroTscalar}. The space of solutions we have found can be characterized by a curve in the $T/\Lambda$,$B/\Lambda^2$ plane. See figure \ref{F:BTcurve}.

\begin{figure}
\centering
\includegraphics[width=0.45\linewidth]{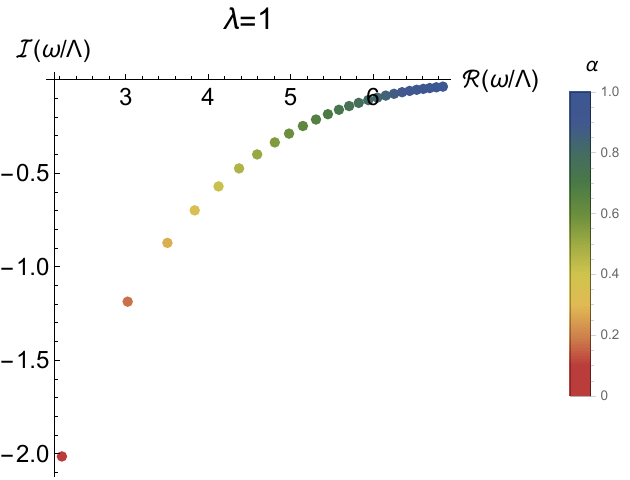}
\hfill
\includegraphics[width=0.45\linewidth]{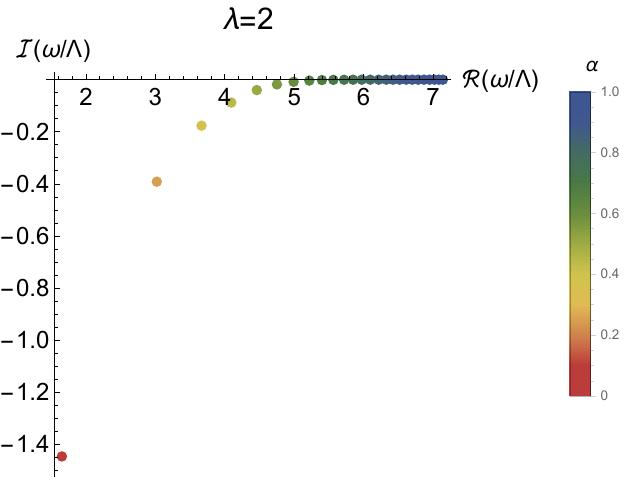}
\caption{Plots of the quasi normal frequencies of magnetically charged black branes along the curve parameterized by $\alpha$ described in figure \ref{F:BTcurve}. As the temperature is lowered the quasi normal modes approach the real axis in agreement with the predictions presented in this work.}
\label{F:QNMfull}
\end{figure}

Given a background solution, we can look for the quasi normal modes associated with it. In figures \ref{F:QNMfull} we plot the real and imaginary components of the quasi normal modes as a function of their location on the $T/\Lambda$,$B/\Lambda^2$ plane. 

 \end{appendix}

\bibliographystyle{JHEP}
\bibliography{LLM}

\end{document}